\documentclass[showpacs,preprintnumbers,prd,superscriptaddress,nofootinbib, twocolumn]{revtex4}

\usepackage{amsmath}
\usepackage{amsfonts}
\usepackage{amssymb}
\usepackage{graphicx}
\usepackage[titletoc]{appendix}
\usepackage{color}
\usepackage[rightcaption]{sidecap}
\usepackage{subfigure}

\DeclareMathOperator{\arcsinh}{arcsinh}

\usepackage{dcolumn}
\usepackage{units}
\usepackage{array}
\usepackage{ctable}
\usepackage{multirow}
\usepackage{siunitx}
\usepackage{longtable}
\usepackage{tabularx}
\usepackage{booktabs}

\usepackage[pdftex,bookmarks=true,colorlinks=true,linkcolor=blue,citecolor=blue]{hyperref}

\graphicspath{{Graphics/}}

\def\be{\begin{equation}}
\def\ee{\end{equation}}
\def\bea{\begin{eqnarray}}
\def\eea{\end{eqnarray}}

\begin{document}

\title{Repulsive regions in  Lema{\^i}tre-Tolman-Bondi  gravitational collapse}


\author{Roberto Giamb\`o}
\email{roberto.giambo@unicam.it}
\affiliation{Scuola di Scienze e Tecnologie, Universit\`a di Camerino, 62032 Camerino, Italy.}
\affiliation{Istituto Nazionale di Fisica Nucleare (INFN), Sezione di Perugia, 06123-Perugia, Italy.}

\author{Orlando Luongo}
\email{orlando.luongo@lnf.infn.it}
\affiliation{Scuola di Scienze e Tecnologie, Universit\`a di Camerino, 62032 Camerino, Italy.}
\affiliation{Istituto Nazionale di Fisica Nucleare (INFN), Laboratori Nazionali di Frascati, 00044 Frascati, Italy.}
\affiliation{NNLOT, Al-Farabi Kazakh National University, Al-Farabi av. 71, 050040 Almaty, Kazakhstan.}

\author{Hernando Quevedo}
\email{quevedo@nucleares.unam.mx}
\affiliation{Instituto de Ciencias Nucleares, Universidad Nacional Aut\'onoma de M\'exico, AP 70543, Ciudad de M\'exico, 04510, Mexico.}
\affiliation{Dipartimento di Fisica and ICRANet, Universit\`a di Roma "Sapienza", I-00185, Roma, Italy.}
\affiliation{Department of Theoretical and Nuclear Physics, Kazakh National University, Almaty 050040, Kazakhstan.}

\date{\today}

\begin{abstract}
We show that in the inhomogeneous Lema{\^i}tre-Tolman-Bondi space-time there are specific regions in which repulsive gravity exists. To find these regions, we use an invariant definition of repulsive gravity based upon the behavior of the curvature eigenvalues. In addition, we analyze the effects of repulsive gravity on the dynamics of the gravitational collapse. In particular, we investigate the collapse in the case of the parabolic solution for the effective scale factor  of the Lema{\^i}tre-Tolman-Bondi metric, corresponding to the marginally bound case. Exploring the corresponding cut-offs at which gravity becomes repulsive, we notice that black holes with dominant repulsive effects are not excluded \emph{a priori}. Indeed, we demonstrate that the collapse leads, in general, to the formation of a central naked singularity; however, for particular values of the free parameters entering the model, black holes with dominant repulsive gravity can exist. We show that the expected physical process is not modified as the marginally bound condition is dropped out. Moreover, we show that this is true independently of the hypothesis that the energy-momentum tensor is built up in terms of pressureless matter. Further, we demonstrate that geodesic deviations  can depend  on  the  sign  of  the  curvature  eigenvalues.
Finally, we give an astrophysical interpretation of black holes with dominant repulsive gravity. Indeed, we argue that compact objects with dominant repulsive gravity could be interpreted as progenitors of Gamma Ray Bursts.
\end{abstract}

\maketitle


\section{Introduction}
\label{uno}

Einstein's general relativity describes the gravitational collapse and predicts the existence of a particular case of space-time singularities, dubbed \emph{black holes}.
Recent observations  represent a direct proof for the existence of such objects \cite{FM}. Moreover, some theoretical studies in general relativity indicate that besides black hole solutions naked singularities may exist \cite{mala1,mala2}.
In particular, a notable property of naked singularities is that they appear as soon as the black hole parameters violate the physical requirements for the existence of an event horizon. But they can also exist under quite general assumptions as exact solutions of the corresponding field equations \cite{primaref}. Consequently, naked singularity solutions could exist even when no black hole counterpart exists \cite{solutions}. Thus, it is impossible to observe a singularity from outside the horizon \cite{penrose}.

Despite rigorous studies\footnote{Many attempts have been made to prove the cosmic censorship hypothesis; see, for example, \cite{hawking}.}, no scenarios of gravitational collapse corroborate the correctness of the conjecture and so there exists also the possibility that, under particular conditions, naked singularities can appear during the evolution of a mass distribution into a gravitational collapse \cite{naked}. For example, it is believed that naked singularities form more often if the collapse is very fast and when the so-formed compact objects are not spherically symmetric. An intriguing result indicates that in an inhomogeneous collapse,  a critical degree of inhomogeneity exists below which black holes form \cite{puglia}. Moreover, naked singularities appear if the degree of inhomogeneity is bigger than the critical value \cite{citazione1}. In conclusion, the collapse speed, the shape of the collapsing object and even inhomogeneities are important factors for determining the final state of a collapse \cite{defy}. It follows that the study of inhomogeneous systems is  relevant for understanding the physics of naked singularities.

Another important aspect of naked singularities is the presence of repulsive gravity, as has been shown for Schwarzschild\footnote{The spherical case is essentially the simplest one. Indeed, one can show that in this case the naked singularity is generated by an effective negative mass situated at the origin of coordinates.}, Kerr and Kerr-Newman space-times \cite{lq12}. Moreover, during the past years,
attempts have been made to investigate the physical effects due to possible regions of repulsive gravity in the case of the homogeneous and isotropic Friedmann-Lema{\^i}tre-Robertson-Walker  space-time, to solve the thorny issue of the observed cosmic speed up \cite{ecx1,ecx2,ecx2bis}.
In particular, in the Friedmann-Lema{\^i}tre-Robertson-Walker  universe the repulsive action of dark energy has been re-framed in terms of repulsive gravity, counterbalancing the action of attractive gravity. This alternative to dark energy, which does not imply a modification of Einstein's gravity \cite{ecx2tris}, is currently a different point of view for describing the effects of the observed cosmic speed up. In other words,
the dark energy dynamical problem is reviewed in terms of a physical mechanism inside the Einstein equations themselves, providing a \emph{geometrical source term}, in which one imagines that, under certain circumstances, the geometry could correspond to a repulsive field\footnote{The most popular mechanisms for repulsive gravity consist in 1) characterizing dark energy by exotic fluids \cite{ecx4,ecx4bis,ecx4tris} or 2) considering extensions and/or modifications of Einstein's gravity \cite{ecx5}. Hence, this mechanism can be considered as a robust \emph{third way} of handling the Universe dynamics, different from the aforementioned ones.}. In view of the aforementioned considerations, one can expect that cosmological inhomogeneities could also be a scenario for the study of naked singularities.

In this work we focus our attention on the spherically symmetric Lema{\^i}tre-Tolman-Bondi (LTB) space-time and investigate the effects of repulsive gravity that emerge from the space-time itself. Our results indicate that the probability of existence of naked singularities cannot be neglected {\it a priori}, if one defines repulsive gravity in terms of invariants of the curvature tensor \cite{def84,def89}. To do so, we present in this work the approach based upon an invariant representation\footnote{This definition can be applied to different types of naked singularities. The physics of each application is reasonable and turns out to give hints toward the physical properties of repulsion.} of the curvature tensor and its eigenvalues \cite{lq12}, where  repulsive gravity is defined in an invariant way by considering the behavior of the curvature tensor eigenvalues. We evaluate the extremal points of the eigenvalues and show which regions indicate a change in the behavior of gravity. We show that the repulsion region is always located at a very short distance from the central gravity source. In other words, we overcome the wide number of intuitive approaches towards  repulsive gravity by means of our invariant definition and we show the compatibility of these results in view of previous outcomes. We explore both the marginally and non-marginally bounds and we highlight the necessary condition for having a naked singularity. We also investigate the  geodesic  deviation  equations  (GDEs) in terms of the curvature eigenvalues. We thus find how gravity becomes repulsive directly from the GDEs as a consequence of the eigenvalue change of sign. We demonstrate that using the GDEs, it is possible to describe repulsion regions that are compatible with those obtained from our geometrically-invariant procedure. Afterwards, we show a critical case that can be interpreted as a black hole with dominant repulsive effects. We interpret this case in the framework of astrophysics and argue that the physical formation of Gamma Ray Bursts may be well explained in this scenario.

The paper is organized as follows. In Sec. \ref{due}, we review the invariant approach which is used to determine the regions in space-time where repulsive gravity can exist. The method is based upon the analysis of the behavior of the curvature eigenvalues. In Sec. \ref{tre}, we present the main features of the gravitational collapse in the inhomogeneous LTB space-time. Sec. \ref{sec:sing} is dedicated to the analysis of the conditions under which the gravitational collapse leads to the formation of naked singularities. Moreover, we find all the repulsive regions of the LTB space-time and compare their locations with the conditions for the formation of singularities in Sec. \ref{cinque}. Section \ref{sec:GDE} is devoted to studying the GDEs in our model, linking changes in the qualitative behavior of their solutions to repulsive gravity effects. In Sec. \ref{sei}, we propose an interpretation of our results in the framework of astrophysics.  We explore the possibility that the formation of ultra-energetic objects, such as Gamma Ray Bursts, can be explained within  our theoretical model. Finally, in Sec. \ref{sec:con}, we summarize and comment our results.

\section{An invariant approach to repulsive gravity with eigenvalues}
\label{due}

Naked singularities have been shown to exist under generic assumptions in general relativity, corresponding to  exact solutions. Every black hole solution possesses a corresponding naked singularity counterpart, emerging as black hole parameters violate the condition for event horizon's existence. The opposite case is not true at all. So, it could happen that naked singularities exist with no black hole counterparts. Thus, it is natural to wonder whether naked singularities describe physical configurations truly existing in Nature \cite{hern,altero1,altero2,altero3}.

Mathematically speaking, with no experimental proofs so far, a limitation is provided by the cosmic censorship conjecture \cite{censura1,censura2}. Other studies indicate that singularities can appear during the evolution of a mass distribution into a gravitational collapse. For example, naked singularities can appear if the degree of inhomogeneity is larger than a given value; moreover, the frequency of naked singularities formation is bigger if the collapse occurs very rapidly and the object breaks down the spherical symmetry somehow. Since these results show that naked singularities can exist \cite{1900,1901,1902,1903}, the physical effects around naked singularities become of  interest in view of  modern observations. In particular, possible regions of repulsive gravity can exist and the need of characterizing them through an invariant definition is essential to predict effects from their existence. Phrasing it differently, this can lead to a proof for the existence of  regions where repulsive gravity can become dominant.

An invariant definition of repulsive gravity for a given metric has been recently suggested, making use of the curvature tensor eigenvalues \cite{2800,2800bis}.  The results obtained so far are physically meaningful because the invariant character of the definition, which is based upon the use of and orthonormal  frame $ \vartheta^a$ and the formalism of differential forms. The orthonormal frame is  the simplest choice for an observer to perform local time, space and gravity measurements. In so doing, all the quantities associated with this frame become coordinate independent. The corresponding orthonormal tetrad is determined by the relationships
\begin{equation}\label{fn}
ds^2 = g_{\mu\nu} dx^\mu dx^\nu= \eta_{ab}\vartheta^a\otimes\vartheta^b\,,
\end{equation}
\vspace{0.2cm}

\noindent with $\eta_{ab}={\rm diag}(-1,1,1,1)$, and $\vartheta^a = e^a_{\ \mu}dx^\mu$.

To compute the Riemann curvature components, using this frame, we employ the first and second Cartan equations,
\be
d\vartheta^a = - \omega^a_{\ b }\wedge \vartheta^b\ ,\
\Omega^a_{\ b} = d\omega^a_{\ b} + \omega^a_{ \ c} \wedge \omega^c_{\ b} = \frac{1}{2} R^a_{\ bcd} \vartheta^c\wedge\vartheta^d\ ,
\ee
whereas for the analysis of the eigenvalues we can consider the bivector representation of the Riemann tensor. This permits us to immediately get its irreducible representation with respect to the Lorentz group. Using the notations and conventions according to which each bivector index
$A,B=1,\ldots,6$  corresponds to two tetrad indices  $A\rightarrow ab$,
i.e., ${\bf R}_{AB}\rightarrow R_{abcd}$  with
\begin{displaymath}
1\rightarrow 01,\ 2\rightarrow 02,\ 3\rightarrow 03,\  4 \rightarrow 23, \  5\rightarrow 31, \  6\rightarrow 12,
\end{displaymath}
the curvature tensor can  be expressed through a $(6\times 6)-$matrix
\cite{gq19}. In particular, all the irreducible components of the Riemann tensor can be written via a bivector representation by $
R_{AB} = W_{AB} + E_{AB} + S_{AB}$, with\footnote{This is the so-called $SO(3,C)$-representation of the Riemann tensor.}
\begin{subequations}
\begin{align}
W_{AB}&=\sigma_3\bf{M}+\sigma_1\bf{N}\,,\\
       E_{AB}&=\sigma_3\bf{P}+\sigma_1\bf{Q}\\
       S_{AB}&=-\frac{\bf{R}}{12}\sigma_3\,,
\end{align}
\end{subequations}
where $\sigma_{1;3}=\overrightarrow{\sigma}_{1;3}\mathbb{I}$, with $\overrightarrow{\sigma}_{1;3}$ the well-known Pauli matrices and $\mathbb{I}$ the identity matrix. Recasting this representation by means of (3$\times$3)-matrices, we can write

\be \label{eq: CurvatureTensor}
{\bf R}_{AB}=\left(
\begin{array}{cc}
	{\bf M}_1 & {\bf L} \\
	{\bf L} & {\bf M}_2 \\
\end{array}
\right),
\ee
where
$$
{\bf L} =\left(
\begin{array}{ccc}
{\bf R}_{14}  & 	{\bf R}_{15}  & {\bf R}_{16} \\
{\bf R}_{15} - \kappa T_{03} &  {\bf R}_{25} &  {\bf R}_{26}  \\
{\bf R}_{16} + \kappa T_{02}  &   \quad {\bf R}_{26}  - \kappa  T_{01} & \quad - {\bf R}_{14} -	{\bf R}_{25}   \\
\end{array}
\right),
$$
${\bf M}_1$ and  ${\bf M}_2$ are $3\times 3$ symmetric matrices
$$
{\bf M}_1=\left(
\begin{array}{ccc}
{\bf R}_{11} &  \quad	{\bf R}_{12} & {\bf R}_{13} \\
{\bf R}_{12} & \quad {\bf R}_{22} &  {\bf R}_{23} \\
{\bf R}_{13} & \quad   {\bf R}_{23} &  \quad - {\bf R}_{11}   -	{\bf R}_{22}  {+} \kappa \left(\frac{T}{2} +T_{00}\right)  \\
\end{array}
\right),
$$
\noindent and finally

\begin{widetext}
$$
{\bf M}_2 =
{   \left( \\
	\begin{array}{ccc}
	-{\bf R}_{11} + \kappa \left(\frac{T}{2} +T_{00}-T_{11} \right)   &  {-} { \bf R}_{12} - \kappa T_{12}   & - {\bf R}_{13} - \kappa T_{13} \\
	{-} { \bf R}_{12} - \kappa T_{12}   &   -{\bf R}_{22} + \kappa \left(\frac{T}{2} +T_{00}-T_{22} \right)   &  - {\bf R}_{23} - \kappa T_{23} \\
	- {\bf R}_{13} - \kappa T_{13}    &    - {\bf R}_{23} - \kappa T_{23}   &   {\bf R}_{11} +	{\bf R}_{22}  {-} \kappa T_{33}   \\
	\end{array}
	\right)   },
$$
\end{widetext}

\noindent with $T=\eta^{ab}T_{ab}$.
This is the most general form {of} a curvature tensor that satisfies Einstein's equations with an arbitrary energy-momentum tensor, where $\kappa\equiv\frac{8\pi G}{c^4}$. Moreover, the traces of the matrices satisfy the relationships
\begin{eqnarray}
&{\rm Tr}({\bf M_1}) &= \kappa \left(\frac{T}{2} +T_{00}\right) \ , \\ &{\rm Tr}({\bf M_2}) &=  \kappa T_{00}\,\\
&{\rm Tr}({\bf R_{AB}}) &= \kappa \left(\frac{T}{2} +2 T_{00}\right)\ .
\end{eqnarray}
All the physical information about  curvature is contained in the eigenvalues $\lambda_i,\ i=1,\ldots 6$ of the matrix ${\bf R}_{AB}$. Particularly, this approach is performed in an invariant way since the eigenvalues behave as scalars under coordinate transformations. It may happen that the sign of at least one eigenvalue changes. In such a case, we interpret this behavior as due to the presence of regions of repulsive gravity. In the same manner, if the gravitational field is finite at infinity, the eigenvalue must have an extremal at some point before it changes its sign and, therefore, the extremal point can be interpreted as the place of \emph{repulsion onset}. Correspondingly, the zeros of the eigenvalues determine the repulsive regions.

\section{Approaching the  Lema{\^i}tre-Tolman-Bondi space-time}
\label{tre}

 We want to consider a spherically symmetric dust cloud collapsing to a singularity in the inhomogeneous approach of the LTB metric. To do so, we can start by solving Einstein's equations,  $R_{\mu\nu}-\frac{1}{2}g_{\mu\nu}R=\kappa T_{\mu\nu}$, for a general spherical metric, written in co-moving coordinates as
\begin{equation}\label{eq:metric}
\mathrm ds^2=-e^{2\nu(r,t)}\,\mathrm dt^2+e^{2\lambda(r,t)}\,\mathrm{d}r^2+R(r,t)^2\,\mathrm d\Omega^2\,,
\end{equation}
and, under the assumption that $T=-\epsilon(r,t)\,\mathrm dt\otimes\frac{\partial}{\partial t}$, we get
\begin{equation}\label{eq:LTB}
\mathrm ds^2=-\mathrm dt^2+\frac{R'(r,t)^2}{1+f(r)}\,\mathrm dr^2+R(r,t)^2\,\mathrm d\Omega^2\,,
\end{equation}
where prime denotes the partial derivative with respect to $r$. In addition, we notice that the function $R(r,t)$ satisfies the evolution equation
\begin{equation}\label{eq:evol}
\dot R=-\sqrt{\frac{2F(r)}{R}+f(r)}\,,
\end{equation}
in which we take the negative root since we are interested in studying the collapse mechanism. The details of calculations can be found in Appendix A. Here, dots denote partial derivative with respect to time, $t$.
One of the two arbitrary functions $F(r)$ is the well established \textit{Misner-Sharp mass} $m=\frac R2 (1-g^{\mu\nu}R,_{\mu}R,_{\nu})$ that plays the role of generalizing the horizon. For the metric \eqref{eq:metric}, it generally reads
\begin{equation}\label{mtmass}
m=\frac R2\left(1-R'^2\,e^{-2\lambda}+\dot R^2
e^{-2\nu}\right)\,,
\end{equation}
and, by virtue of the second relation of Eqs. \eqref{equazioniappendice} of Appendix A, it turns out to be\footnote{Notice that this happens as a consequence of the fact that we are considering dust, i.e. matter with vanishing pressure. Without this condition one gets a modified  Misner-Thorne mass, leading to a $m=F(r,t)$ function, when radial and transverse pressures are included into the LTB metric.} $m=F(r)$ as in Eq. \eqref{eq:evol}.

The other function, namely $f(r)$, is the so--called \textit{velocity function} \cite{Joshi2007}, since it is related to the velocity of the dust cloud at initial time, once the mass profile $F(r)$ and an initial condition on $R$  are fixed. Setting $f$ too, one solves Eq. \eqref{eq:evol} and finds the unknown function $R(r,t)$, representing the radius of the shell $r$ at comoving time $t$. The corresponding analysis of Eq. \eqref{eq:evol} leads to two main cases, the first, $f(r)=0$, is  commonly known as the \textit{marginally bound}, and will be treated in full details in the next Sections. The so called \textit{bound} (resp. \textit{unbound}) case, corresponding to $f(r)<0$ (resp. $f(r)>0$), is more involved since an implicit  resolution of Eq. \eqref{eq:evol} is invoked, and will be treated in Section \ref{sec:nonmarginally}.

Therefore, let us consider $f(r)=0$. Setting $R=r$ at $t=0$ we can find the following explicit form for $R(r,t)$:
\begin{equation}\label{eq:R}
R(r,t)=r[1-k(r) t]^{\frac{2}{3}},\quad\text{\ where\ }\quad k(r)=\frac32\sqrt{\frac{2F(r)}{r^3}}.
\end{equation}
In this case, the solution is completely determined by the choice of the mass $F(r)$, that in the following will be taken  in the class $\mathcal C^\infty([0,r_b])$ for some $r_b>0$. The value $r_b$ represents the comoving boundary of the star: Moreover, an external matching with the Schwarzschild solution of mass $M=F(r_b)$ at the junction hypersurface $\{r=r_b\}$ is possible, since the radial pressure vanishes along this hypersurface\footnote{For the sake of completeness, it identically vanishes on all the internal space-time.} \cite{Magli1997}.

We will also suppose that the mass $F(r)$ is a non negative function such that the initial energy $\epsilon_0(r):=\epsilon(r,0)$  is a decreasing function of $r$, which continuously extends until the central shell $r=0$ in order for the solution to be completely regular at $t=0$. Using Einstein equations we obtain
\begin{equation}\label{eq:e}
\epsilon(r,t)=\frac{F'(r)}{4\pi R^2 (r,t) R'(r,t)}
\end{equation}
and then by virtue of Eq. \eqref{eq:R}, we immediately get
\begin{equation}\label{epsy}
\epsilon_0(r)\equiv\epsilon(r,0)=\frac{F'(r)}{4\pi r^2}\,.
\end{equation}

To guarantee that Eq. \eqref{epsy} is mathematically well-defined in $r=0$, it must necessarily be
\begin{equation}\label{eq:Fexp}
F(r)=F_0 r^3+F_n r^{3+n}+\mathcal O(r^{4+n})\,,
\end{equation}
with $n$ a positive integer number that can be chosen arbitrarily. Moreover, since the Misner-Thorne mass, Eq. \eqref{mtmass} might be positive definite, we should have
\begin{equation}\label{eq:Fn}
n>0,\qquad F_0>0,\qquad F_{n}<0,
\end{equation}
which guarantee that $F(r)>0$ and $\epsilon^\prime(0,r)<0$. Furthermore, the condition $F_n<0$ guarantees that, departing from the center, the mass decreases.

Choosing, up to scaling, $F_0=\tfrac29$ and defining the positive parameter $a:=-\tfrac94 F_n$, we have from \eqref{eq:R} and \eqref{eq:Fexp} that $k(r)$ is normalized as $k(0)=1$ and reads:
\begin{equation}\label{eq:k}
k(r)=1-ar^n+\mathcal O(r^{n+1}).
\end{equation}
In the following we will see that fixing the positive parameter $a$ and the integer $n$ determines the collapse endstate of the dust cloud.

\section{Singularity formation}
\label{sec:sing}

From Eq. \eqref{eq:e}, we see that a singularity occurs either for $R=0$ or when $R'=0$. Recalling Eq. \eqref{eq:R}, we see that the first case corresponds to either $r=0$ or $t=1/k(r)$. In principle, the region $r=0$ is not a true singularity for all collapsing times $t$, as one can see by obtaining that

\begin{equation}\label{5}
\epsilon(0,t)=\frac23(1-t)^{-2}\,,
\end{equation}
and, therefore, the central shell becomes singular at time $t=1$, corresponding to the value of the singular function for $r=0$:

\begin{equation}\label{6}
t_s(r)=\frac{1}{k(r)}=1+a r^n+\mathcal O(r^{n+1})\,
\end{equation}
that marks, for each shell $r\in[0,r_b]$, the time of collapse.

\noindent The boundary region where $R=0$ is called a \textit{shell focusing singularity}, but we must first exclude that $R'$ does not vanish along the evolution, prior to $R=0$, i.e. that a \textit{ shell crossing singularity} forms before the shell is focusing. With cumbersome algebra, one obtains that $R'=0$ if $t=t_{sc}(r)$, where
\begin{equation}\label{7}
t_{sc}(r)=\frac{1}{k(r)+\tfrac23 r k'(r)}\,.
\end{equation}
In the following we presume $k'(r)< 0$ as $r>0$ (which is compatible with the sign choices made in \eqref{eq:Fn}, due to \eqref{eq:k}), in order to have  $t_{sc}(r)>t_s(r)$ and then no shells crossing singularity appear. Hence, $t=t_s(r)$ represents the actual future boundary of the space-time. Note that the denominator in the righthand side of \eqref{7} equals $\tfrac{3\epsilon_0(r)}{2k(r)}$ and so it is a strictly positive quantity. This fact will be useful in later applications. .

To study the collapse endstate, it is worthwhile recalling that the aforementioned assumptions exclude that the noncentral singularities $\{(t_s(r),r)\,:\,r>0\}$ are naked \cite{GGM2002}.
That notwithstanding, one finds that the apparent horizon, implicitly defined by $R=2m$, is described by a curve in the $(r,t)$--plane given by
\begin{equation}\label{eq:th}
t_h(r)=\frac{1}{k(r)}\left(1-\frac{8}{27}k(r)^3 r^3\right)=t_s(r)-\frac8{27}k(r)^2 r^3,
\end{equation}
and, therefore, the central shell $r=0$ becomes trapped at time $t_h(r)=1$, i.e. the same co-moving time it collapses. Thus, it is possible that the null outgoing geodesics, emitted from the central singularity $(r,t)=(0,1)$, may exist in the region $I\!I=\{t<t_h(r)\}$, giving rise to a (locally) naked singularity. It is a consolidate result \cite{EardSmarr1979,Joshi2007,NolanMena2001} that the parameters $a$ and $n$ defined in \eqref{eq:k} dictate this possibility. In particular, it has been shown that \textit{the singularity is naked if $n=1,2$ and $n=3$ but in this critical case it must be}

\begin{equation}\label{critical}
a>a_c:=\tfrac{26+15\sqrt{3}}{3}F_0\simeq     3.85\,.
\end{equation}

\noindent In the other cases, the central singularity is covered by the horizon as well as the non-central ones.  The proof is obtained by a careful study of the null geodesic equation asymptotically near $r=0$, and it is remarkable that the above picture remains qualitatively unaffected for a larger class of models, including LTB models where a nonzero arbitrary function $f(r)$ is taken\footnote{See \cite{GGMP2002,GGMP2003} for details.}.

\begin{figure*}
\centering
\includegraphics[width=5in]{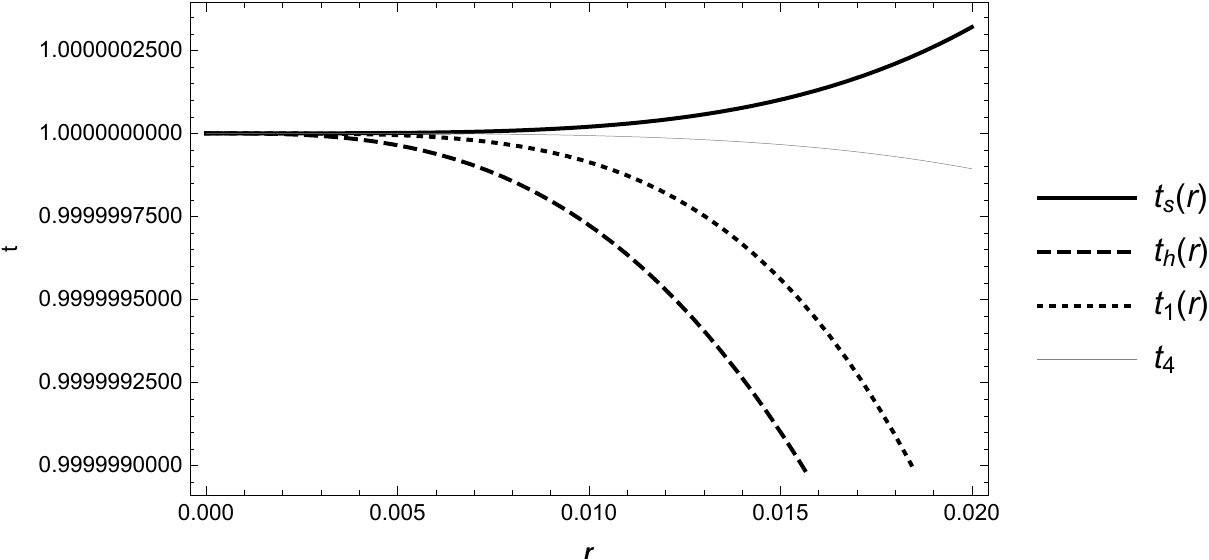}
\caption{Plots of various $t$'s functions for arbitrary values of $n$ and $a$. We choose $n=3$ and $a=\frac{5}{27}$ and we plot the singular curve, i.e. $t_s\equiv\frac{1}{k}$, the horizon curve, i.e. $t_h\equiv t_s\left(1-{8\over27}k^3r^3\right)$, and $t_1$ and $t_4$. It is possible to notice that, for different values of $n$, the shapes are not intertwined among them up to a given radius.}\label{m1}
\end{figure*}

\begin{figure*}
\centering
\begin{tabular}{lr}
\includegraphics[width=3.6in]{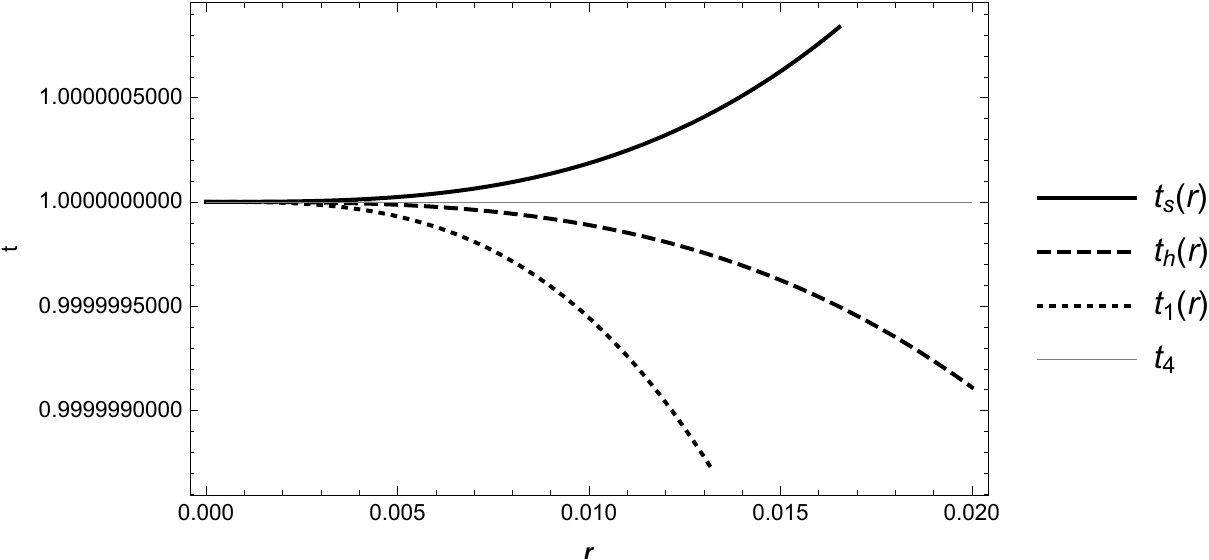} &
\includegraphics[width=3.6in]{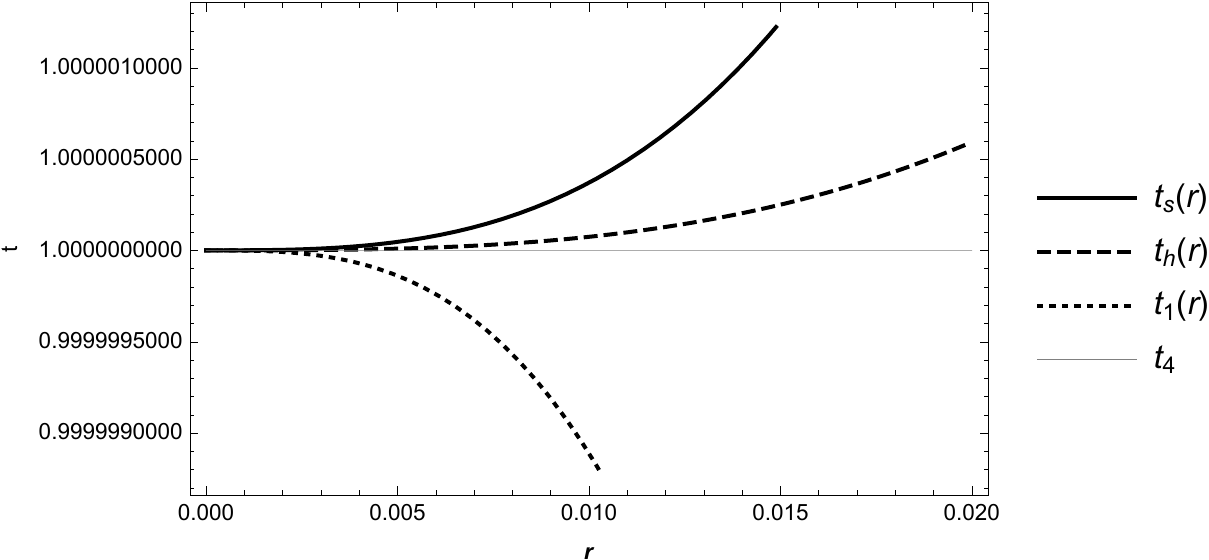}
\end{tabular}
\caption{Plots of various $t$'s functions for arbitrary values of $n$ and $a$. We choose $n=3$ for both the curves and two different values of $a$, i.e. on the left $a={5\over27}$ and the right $a={10\over27}$. We plot the singular curve, i.e. $t_s\equiv\frac{1}{k}$, the horizon curve, i.e. $t_h\equiv t_s\left(1-{8\over27}k^3r^3\right)$, and $t_1$ and $t_4$. One can notice that, for different values of $n$, the shapes are not intertwined among them up to a given radius. Strong departures are found for $t_h$ that turns out to be the most sensible curve to $a$.}\label{m2}
\end{figure*}

\section{Repulsive regions}\label{cinque}

To investigate the repulsion regions in the LTB metric, we follow the approach outlined in Sec. \ref{due}. In the parabolic case, the orthonormal tetrad can  be chosen as
\begin{equation}
\vartheta^0 = dt\ , \quad \vartheta^1 =  R^\prime dr\ ,\quad \vartheta^2 = R d\theta\ ,\quad \vartheta^3 = R  \sin\theta d\varphi\ .
\end{equation}
The computation of the corresponding curvature matrix is straightforward and the analysis of its  eigenvalues leads to
\begin{align}\label{eq:lambda}
\lambda_1&=\frac{\ddot R'}{R'},\qquad\qquad \lambda_2=\frac{\dot R^2-f}{R^2},\\
\lambda_3&=\lambda_5=\frac{\ddot R}{R},\qquad \lambda_4=\lambda_6=\frac{\dot R\dot R' - \tfrac12 f'}{R R'}\,.
\end{align}

According to the definition of repulsive gravity described in Sec. \ref{due}, a change of sign in any eigenvalue indicates the presence of a region in which repulsion dominates over attraction. We thus search for the zeros
of the $\lambda_i$'s belonging to the region $\mathcal J=\{ 0\le t< t_s(r)\}$.

\subsection{Changing signs of eigenvalues}

Using Eq. \eqref{eq:evol}, and recalling that, along the evolution, $R'>0$ to enable shells do not interact to each other and $\dot R<0$ to guarantee the collapse, in $\mathcal J$,  it is easy to get:
\begin{itemize}
\item $\lambda_1$ changes sign where
\begin{equation}\label{eq:t1-gen}
2 F(r) R'(r,t)=F'(r) R(r,t);
\end{equation}
the zeroes are given by the set $(r,t_1(r))$, where
\begin{multline}\label{eq:t1}
t_1(r)=
\frac{1}{k(r)}\left( 1+\frac{4rk'(r)}{3k(r)+2rk'(r)}\right)\\
=1-\frac{4n-3}{3}a r^n+\mathcal O(r^{n+1}).
\end{multline}
Note that, as we have observed before, the quantity $3k(r)+2rk'(r)$ is strictly positive.
\item $\lambda_2>0$ and $\lambda_3=\lambda_5=-\frac{F(r)}{R(r,t)^3}<0$. Both the eigenvalues do not change sign.
\item $\lambda_4=\lambda_6$ change sign where $\dot R'$ does, i.e. where
\begin{equation}\label{eq:t4-gen}
F(r) R'(r,t)=F'(r) R(r,t)
\end{equation}
This happens on the points $(r,t_4(r))$, where
\begin{multline}\label{eq:t4}
t_4(r)=
\frac{1}{k(r)}\left( 1+\frac{rk'(r)}{3k(r)+2rk'(r)}\right)\\
=1-\frac{n-3}{3} a r^n+\mathcal O(r^{n+1})
\end{multline}
\end{itemize}

The shapes of $t_1$ and $t_4$ have been reported in Figs. 1 and 2 for different values of $n$ and $r$, distinguishing the cases $n=1,2,3$ from $n\geq4$. Observe that, from the exact expressions \eqref{eq:t1} and \eqref{eq:t4}, and recalling $k'(r)<0$, we always have $t_1(r)> t_4(r)$, $\forall r\in]0,r_b]$.

\noindent Let us now study the mutual positions of the curves $t_1(r)$ Eq. \eqref{eq:t1}, $t_4(r)$ Eq. \eqref{eq:t4}, $t_h(r)$ Eq. \eqref{eq:th} in the $(r,t)$--plane near the centre $r=0$, as the parameters $n$ and  $a$ defined in Eq. \eqref{eq:k} vary. We remark that at $r=0$ all these curves coincide with each other and with the singular curve $t_s(r)$.
For $r>0$ sufficiently close to $r=0$, it is found that:
\begin{itemize}
\item in  case $n=1,2$, we have $t_1(r)<t_4(r)<t_h(r)$: both $\lambda_1$ and $\lambda_4$ changes sign \textit{before} the shell labeled $r$ gets trapped;
\item in case $n\ge 4$, we have that $t_h(r)<t_1(r)<t_4(r)$: both  $\lambda_1$ and $\lambda_4$ changes sign \textit{after} the shell labeled $r$ gets trapped;
\item in the critical case $n=3$, where both naked singularities and black holes may take place -- depending on the value of $a$, we find that:
\begin{itemize}
\item if $a<\tfrac2{27}$ then $t_h(r)<t_1(r)<t_4(r)$;
\item[\,]  \,
\item if $a\in]\tfrac2{27},\tfrac8{27}[$ then $t_1(r)<t_h(r)<t_4(r)$;
\item[\,]  \,
\item if $a>\tfrac8{27}$ then $t_1(r)<t_4(r)<t_h(r)$;
\end{itemize}
in the -- highly non generic -- transition cases, one should in principle consider higher order terms of $k(r)$. With some calculus one finds that $\lambda_4$ changes sign on the horizon in case $F(r)=\frac32 W\left(\frac{4}{27}r^3\right)$,
where $W(z)$ is the principal branch of Lambert\footnote{Sometimes called ProductLog function.}  $W$--function ($t_4(r)=t_h(r)$). Analogously, the choice of mass $F(r)=\frac{3}{8}W\left(\frac{16}{27}r^3\right)$ corresponds to a situation where $\lambda_1$ exactly changes sign on the horizon ($t_1(r)=t_h(r)$).
\end{itemize}
We can conclude that, in all situations where a central naked singularity occurs, both $\lambda_1$ and $\lambda_4$ changes sign before horizon formation. However, this feature also appears in some situations of the critical case $n=3$, $a<a_c=\tfrac{2(26+15\sqrt{3})}{27}$, where the central singularity  is not naked.

\subsection{Analysis of the $f(r)\ne 0$ case}\label{sec:nonmarginally}

Let us now  drop out the marginally bound hypothesis and to see whether our previous results are preserved or not.

To do so, using again the initial condition $R=r$ at $t=0$, one can integrate Eq. \eqref{eq:evol}, obtaining $t\equiv t(r,R)$, i.e. $t$ as a function of $(r,R)$ \cite{Singh1996}
\begin{equation}\label{eq:t-impl}
    t(r,R)= \left. \sqrt\frac{s^3}{2F(r)}\,\Gamma\left(-\frac{f(r)s}{2F(r)}\right)\right|_{s=R}^{s=r},
\end{equation}
where $\Gamma(y)$ is the function
\[
\Gamma(y)=
\begin{cases}
-\frac{\arcsinh\sqrt{-y}}{(-y)^{3/2}}-\frac{\sqrt{1-y}}{y},&\quad y<0,\\
\,\\
\frac23,&\quad y=0,\\
\,\\
\frac{\arcsin\sqrt{y}}{y^{3/2}}-\frac{\sqrt{1-y}}{y},&\quad 0<y\le 1.
\end{cases}
\]
Thus, concerning this case it is useful to perform the analysis within the so--called {\sl{area radius}} coordinate system $(r,R)$. This region is well-defined in the collapse state, because Eq. \eqref{eq:evol} dictates $\dot R<0$, so that  $R$ plays the role of a sort of 'reverse' comoving time. This analysis has been successfully used in \cite{GGMP2002,GGMP2003} to find a broader class of collapsing solutions whose endstate -- black hole vs naked singularity -- has been classified in terms of the Taylor expansion of some functions, thus including the present model as a special case.

To this aim, we can immediately observe  that requirements similar to those made in the marginally bound case -- see Eq. \eqref{eq:e} and following discussion -- lead to the assumption that $F(r)$ is   given again by Eq. \eqref{eq:Fexp}. In addition, we have
\begin{equation}
    f(r)=f_0 r^2+ f_{\tilde{m}} r^{2+\tilde{m}}+\mathcal{O}(r^{3+\tilde{m}}),
\end{equation}
where $\tilde{m}$ is a positive integer. It will be also  necessary to consider the function
\begin{equation}\label{eq:H}
    H(r,R)=2F(r)+R f(r),
\end{equation}
that by construction contains third order leading terms, whereas the following leading term is of order $N=\min\{n,\tilde{m}+1\}$. As proved in \cite{GGMP2002,GGMP2003}, the central singularity is naked if $N=1,2$ or $N=3$ but it must be
\begin{equation}\label{eq:a-nonmarg}
    a:=-\int_0^1\frac{(2F_3+f_3\tau)\sqrt\tau}{2(2F_0+f_0\tau)^{3/2}}\,\mathrm d\tau>\frac{2F_0}{3}\xi_c,
\end{equation}
where $\xi_c=\frac{26+15\sqrt{3}}{2}$. It is straightforward to see that Eq. \eqref{eq:a-nonmarg} reduces to Eq. \eqref{critical}, in the previously discussed marginally bound case. The asymptotic behavior near the centre of the apparent horizon turns out to be $R_h(r)=2F_0 r^3+\mathcal{O}(r^4)$ as before.

Thus, to evaluate repulsive gravity effects, we first observe that again the only significant cases are given by $\lambda_1$ and $\lambda_4=\lambda_6$, and the equations where a sign change occurs are given again by \eqref{eq:t1-gen} and \eqref{eq:t4-gen}, respectively.

To translating these equations in area radius coordinates, let us compute the curves $R_1(r)$ and $R_4(r)$ where the eigenvalues change sign. With some calculations one finds that the leading terms of this curves is of order $r^{(2N+3)/3}$. We conclude that, when $N=1,2$, given $r$ sufficiently small, both $R_1(r)$ and $R_4(r)$ are larger that $R_h(r)$. As a consequence, during the collapse both eigenvalues change their signs before the apparent horizon. In the critical case $N=3$ we have $R_j(r)\cong R_{j0}r^3$, $j\in\{1,4\}$, with
$$
R_{j0}=\left(\frac{6a}{j}\sqrt{2F_0}\right)^{2/3}\,,
$$
and when the singularity is naked, using Eq.  \eqref{eq:a-nonmarg}, we see that $R_{j0}>2F_0$ and then again both eigenvalues change their signs before the apparent horizon.

Then, we can conclude again that an eigenvalue sign change is \emph{a necessary but not sufficient condition for the formation of a central naked singularity.}

\section{Getting repulsive gravity from geodesic deviation}\label{sec:GDE}

The GDEs describe gravitational effects in terms of the relative acceleration between geodesic motions  \cite{Wald1984,mtw17,omg}, involving the Riemann curvature tensor. Hence, it should be possible to work out the above approach of repulsive gravity even in the framework of GDEs. We can thus confront our technique with GDEs, by rewriting the corresponding equations in terms of the curvature eigenvalues and checking what happens as the eigenvalues change signs.

Let $x^\alpha(\sigma,\tau)$ be a congruence of geodesics in such a way that, for each fixed $\sigma$, the map $\tau\mapsto x^\alpha(\sigma,\tau)$ is a geodesic; in particular, $\gamma(\tau):=x(0,\tau)$ will be called the \textit{reference} geodesic.
The velocity field along each geodesic is given by $u^\alpha(\sigma,\tau)=\partial_\tau x(\sigma,\tau)$, whereas $n^\alpha$ is the vector field linking two nearby geodesics, $n^\alpha(\sigma,\tau)=\partial_\sigma x^\alpha(\sigma,\tau)$. Then $\dot\gamma^\alpha(\tau)=u^\alpha(0,\tau)$, and the deviation from the reference geodesic is described by the vector field $J^\alpha(\tau):=n^\alpha(0,\tau)$. It is well-known that $J^\alpha$ is a Jacobi vector field along $\gamma$, satisfying the GDEs
\begin{equation}\label{eq:Jac}
\ddot J^\alpha(\tau)+R^\alpha_{\ \mu\nu\xi}(\gamma(\tau))\,\dot\gamma^\mu(\tau) J^\nu(\tau)\dot\gamma^\xi(\tau)=0,
\end{equation}
where $R^\alpha_{\ \mu\nu\xi}$ is the Riemann curvature tensor expressed in terms of the Christoffel symbols of the Levi--Civita connection of $g_{\mu\nu}$ as
$R^\alpha_{\ \mu\nu\xi}=\Gamma^\alpha_{\mu\xi,\nu}-\Gamma^\alpha_{\mu\nu,\xi}
+\Gamma^\eta_{\mu\xi}\Gamma^\alpha_{\eta\nu}-
\Gamma^\eta_{\mu\nu}\Gamma^\alpha_{\eta\xi}$.

Let us take a null radial geodesic for the metric \eqref{eq:LTB} as reference geodesic, $\dot\gamma^1=\frac{\sqrt{1+f(\gamma^1)}}{R'(\gamma^0,\gamma^1)}\dot\gamma^0$. Moreover, we can limit ourselves to the case of  a geodesic congruence on the equatorial plane, $x^2=\theta=\pi/2$, in such a way that $n^2$ -- and then $J^2$ --  vanishes.
Then the GDE \eqref{eq:Jac} takes the form (the argument $\tau$ is dropped out)
\begin{subequations}\label{theequations}
\begin{align}
&\ddot J^0+\lambda _1 \left(J^0-\frac{ R'\left(\gamma^0,\gamma^1\right)}{\sqrt{1+f\left(\gamma^1\right)}}J^1\right)(\dot\gamma^0)^2 =0,\\
&\ddot J^1+\lambda _1  \left(\frac{\sqrt{1+f\left(\gamma^1\right)}}
{R'\left(\gamma^0,\gamma^1\right)}J^0-J^1\right)(\dot\gamma^0)^2=0,\\
&\ddot J^3+
\left(\lambda _4-\lambda _3\right)  (\dot\gamma^0)^2 J^3=0\ ,\label{eq:J3}
\end{align}
\end{subequations}
showing that indeed the geodesic deviation can depend on the sign of the curvature eigenvalues. From the above equations, we first observe that the third equation is not affected by a sign change of the eigenvalues, since $\lambda_4-\lambda_3$ holds a definite sign in our model. This should not surprise due to the specificity of the reference geodesic fixed here. It is worthwhile recalling that the same happens for a spherically symmetric space-time with charge. Indeed, in the case of the Reissner-Nordstr\"om space-time, one can find that the counterpart of equation \eqref{eq:J3} is trivially $\ddot J^3=0$. In our case, the spherical symmetry is analogously preserved and we find $\ddot J^3=KJ^3$, where $K$ is a constant that depends upon the fixed sign of $\lambda_4-\lambda_3$. The corresponding exponential solutions are thus easy to get.

A different behavior happens for $\ddot J^0$ and $\ddot J^1$. In this case, it is easy to show that the sign of the eigenvalues influences the dynamics. To show that, we can  perform a qualitative study of the first couple of the above equations. We then approximate them nearby a given point of the reference geodesic and we study the behavior of the Jacobi field near that point. We can assume the first component of four-velocity to be identically one and posing $a\equiv\frac{R^\prime}{\sqrt{1+f(r)}}$, we get
\begin{align}\label{geodesics1}
\ddot J^0 & = -\lambda_1 J^0 + a \lambda_1 J^1\,,\\
\ddot J^1 & = -\frac{\lambda_1}{a} J^0 + \lambda_1 J^1\,,
\end{align}
where we set the following initial conditions $((J^0)'(0), ((J^1)'(0))=(0,0)$. In such a way, we do not observe acceleration effects that are not purely gravitational. In fact, there are no acceleration effects induced by the initial conditions if we impose that the first derivatives vanish as above.

The solutions of Eqs. \eqref{geodesics1} are
\begin{align}
J^0 &=  j^0 + {\lambda_1\over 2}\tau^2\left(j^1a-j^0\right)\,,\\
J^1 &=  j^1 + {\lambda_1\over 2a}\tau^2\left(j^1a-j^0\right)\,,
\end{align}
\noindent where we chose the following initial conditions: $J^0(0)\equiv j^0$ and $J^1(0)\equiv j^1$. Notice that  $v^0\equiv \frac{j^0-aj^1}{2}\neq0$, because otherwise the initial Jacobi vector would be parallel to the initial velocity of the geodesic and, therefore, the variation would be entirely lying on the reference geodesic. Consequently, we can recast our equations with the variables $u={1\over 2} (t-a r), v={1\over 2} (t+a r)$. Then, we obtain

\begin{align}
   {1\over2} (J^0(\tau) - a J^1(\tau)) &= v^0\,,\\
   {1\over2} (J^0(\tau) + a J^1(\tau)) &= w^0 - \tau^2  v^0\lambda_1\,,
\end{align}
where $w^0\equiv j^0+a j^1$ has a definite sign.

Now we can observe that the norm of the above vector increases or decreases, depending on the sign of $\lambda_1$. This is interpreted as follows: the sign of the eigenvalue influences the dynamics of the Jacobi field $J$, indicating how gravity acts over the dynamics itself. This is physically due to the fact that gravity changes its sign and confirms the goodness of the approach that makes use of eigenvalues. A more general approach can be performed as one extends the previous calculation to a non-radial case. There, a similar result can be obtained, indicating again that the repulsive character of gravity depends on the sign of the curvature eigenvalues.

\section{Theoretical discussion on compact object formation}\label{sei}

As mentioned above, the change of eigenvalue signs, leading to the presence of repulsive regions \emph{before the horizon forms}, turns out to be \emph{a necessary condition to have a naked singularity, albeit not sufficient}. Consequently, to enable the formation of  a naked singularity, one might admit the existence of further physical mechanisms entering the above puzzle, with the property of \emph{delaying the horizon formation}.

Further, if gravity becomes repulsive, the corresponding  eigenvalues change their behavior, i.e. showing at which surface $(r,t)$ repulsive gravity becomes dominant. However, the region where the eigenvalue vanishes implies that the eigenvalue provides an extremal at some hyperplane before it changes its sign, so that the repulsion region is \emph{the first extremal that appears in a curvature eigenvalue as we approach the origin of coordinates from infinity}. This indicates that the onset of repulsion occurs if \emph{only one of the eigenvalues changes sign}, independently from the rest.

In other words, the mechanism  of the repulsive onset does not affect the formation of a naked singularity with repulsive gravity and \emph{is not against the formation of black holes in which the repulsive effects are dominant}. In view of previous investigations \cite{lq12,LuoQue2014}, we notice that the only possibility, to allow black holes with dominating repulsive interactions, is to take into account the inhomogeneity of the LTB metric.

So, without adding any further mechanism that selects naked singularities instead of black holes, a possible interpretation of the case $n=3$ is that the LTB metric is a  \emph{source} for the creation of highly emitting compact objects, subject to an explosion due to their ``instability". Here, the involved masses, for the cases of black holes and/or naked singularities, are huge. So, the simplest approach, involving the discussed LTB collapse, suggests that Gamma Ray Bursts may form. In such a case, the progenitors are interpreted as LTB black holes dominated by repulsive gravity. Rephrasing it differently, we can say that \emph{Gamma Ray Burst progenitors could be LTB black holes, in which the repulsive effects are dominant}, since inhomogeneities lead somehow to \emph{instabilities} pushing up the black hole to explode\footnote{For the sake of clearness, this speculation involves the LTB case only. A more general demonstration that the rate of inhomogeneities may give hints toward the formation of Gamma ray Burst is beyond the purposes of this paper.}.

\section{Final outlooks and perspectives}
\label{sec:con}

In this work, we investigated the dynamics of the  gravitational collapse in the inhomogeneous LTB space-time and, in particular, the cases in which naked singularities appear. In addition, we explored the possibility to get repulsive gravity in particular regions. We also discussed how repulsive gravity affects the dynamics of the collapse. To do so, we applied an invariant prescription, according to which the presence of repulsive gravity can be detected through the behavior of the curvature tensor eigenvalues. The advantage is that eigenvalues are \emph{scalars with respect to coordinate transformations}.  In fact, we used the intuitive approach that if an eigenvalue changes sign at a given space-time region, it indicates that repulsive gravity dominates over attractive gravity in that region.

On the other hand, an independent way to measure the effects of gravity through curvature is by using the  geodesic deviation. Accordingly, we performed an analysis of the geodesic deviation equations in the LTB space-time and found that indeed the curvature eigenvalues enter the equations explicitly and their sign  affect directly the deviation. This shows that the intuitive approach of using curvature eigenvalues to determine the character of gravity can be verified by using the more physical approach of geodesic deviation.

Thus, we investigated all the zeros of the eigenvalues and compared their locations in space-time with the conditions under which naked singularities appear in the case of  parabolic and non-parabolic LTB space-times.
We found a coincidence between the appearance of naked singularities and repulsion regions, implying that repulsive gravity effects are relevant for studying the dynamic behavior of naked singularities. Analogous results have been found in the context of naked singularities with black hole counterparts \cite{LuoQue2014} and in homogeneous and isotropic cosmological models \cite{LuoQue2018}.

We also found a particular case in which the gravitational collapse, leading to the formation of a black hole, is also accompanied by a change of sign in one eigenvalue. This seems to be a very special   solution in which the parameters are forced to satisfy specific cut-off conditions. Particularly, we got this result more akin to an actual black hole dominated by repulsive gravity. Thus, our interpretation is  that the inhomogeneous LTB metric provides a new physical  case, in which a black hole with dominant repulsive gravity can exist.

We increased complexity working out the $f(r)\neq0$ case. One basic motivation for this choice is that a non-perfectly marginally bound case can occur in general. We found that the outcomes did not go in the opposite direction of our $f(r)=0$ case and, in exchange for the cost of complexity, one gets analogous outcomes as previously found. We therefore recovered LTB black holes dominated by repulsive gravity. Hence, we concluded that \emph{black holes with dominating repulsive effects are a  phenomenon whereby the hypothesis of dust is not essential.} In other words, the picture is extended to matter different from pure dust, in agreement with previous efforts developed in the literature.

Nevertheless, there are no  serious astrophysical proposals for actually detecting black holes with dominating repulsive effects. We therefore pursued a more concrete theoretical goal sketching out physical speculations on compact objects based on this phenomenon. For the sake of clearness, we should stress that the change of signs of eigenvalues is a necessary condition to get repulsive gravity. So, assuming that repulsive effects are present, we apply our  results to astrophysical objects. In so doing, we interpreted Gamma Ray Burst formation as a consequence of LTB inhomogeneities.
Indeed, the here-developed mathematical scheme points out that our results are not a stand-out exception and black holes with dominating repulsive regions may be interpreted as progenitors for highly-energetic compact objects, such as Gamma Ray Bursts. Our treatment has been obtained in the framework of LTB metric only. Future efforts will clarify whether this can happen as genuine signature due to inhomogeneities. Moreover, we will clarify the mechanisms behind the born of Gamma Ray Burst progenitors in view of our results, giving additional examples in the field of astrophysics.

\vspace{0.5cm}

\acknowledgements

O.L. acknowledges the support of INFN (iniziative specifiche MoonLIGHT-2) and warmly thanks Marco Muccino for his thorough comments. The authors warmly thank an anonymous referee for valuable suggestions that helped to improve the quality of the manuscript. The work was supported in part by the Ministry of Education and Science of the Republic of Kazakhstan, Program 'Fundamental and applied studies in related fields of physics of terrestrial, near-earth and atmospheric processes and their practical application' IRN: BR05236494, BR05236322 and AP05133630.
This work was partially supported  by UNAM-DGAPA-PAPIIT, Grant No. 114520, and
Conacyt-Mexico, Grant No. A1-S-31269.



\clearpage

\appendix

\onecolumngrid

\section*{Appendix A}

Einstein's gravity in the case of LTB metric leads to

\begin{subequations}\label{eq:Einstein}
\begin{align}
G^t_t&=-\frac{2m'}{R'R^2}+\frac{2\dot R e^{-2\nu}}{R'R}\left(\dot R'-\dot R\nu'-\dot\lambda R'\right),\label{Gtt}\\
G^r_r&=-\frac{2\dot m}{\dot RR^2}-\frac{2R' e^{-2\lambda}}{\dot RR}\left(\dot R'-\dot R\nu'-\dot\lambda R'\right),\label{Grr}\\
G^t_r&=\frac{2e^{-2\nu}}{R}\left(\dot R'-\dot R\nu'-\dot\lambda R'\right),\label{Gtr}\\
G^r_t&=-\frac{2e^{-2\lambda}}{R}\left(\dot R'-\dot R\nu'-\dot\lambda R'\right),\label{Grt}\\
G^\theta_\theta &= G^\phi_\phi = \frac 1R\left\{
e^{-2\lambda}[(\nu''+\nu'^2-\nu'\lambda')R+R''+R'\nu'-R'\lambda']-\right.\\
&\left.e^{-2\nu}[(\ddot\lambda+\dot\lambda^2-\dot\lambda\dot\nu)R+\ddot
R+\dot R\dot\lambda-\dot R\dot\nu]\right\}.
\end{align}
\end{subequations}

\noindent Solving Einstein's equations \eqref{eq:Einstein} for $\nu,\lambda,R$ and $\epsilon$ and using  \eqref{Gtt}--\eqref{Gtr} we find
\begin{equation}\label{equazioniappendice}
m'=4\pi\epsilon R^2 R',\quad
\dot m=0, \quad
\dot R'=\dot R\nu'+\dot\lambda R'=0.
\end{equation}
So, $
m=F(r)$ and $\nu=\nu(t)$, and by means of $
\frac{\mathrm dT}{\mathrm dt}=e^{\nu(t)}$, we imagine that $e^{\nu(t)}=1$, i.e. $
\nu(t)\equiv 0$, having $\tfrac{\dot R'}{R'}=\dot\lambda$ which is perfectly solvable to give
$
R'=e^{g(r)+\lambda}$, with $g(r)$ the initial setting. Simply imposing all the above conditions, we get $
F(r)=\frac R2\left(1+\dot R^2-e^{2g(r)}\right)$, and solving it  with the position $f(r)=e^{2g(r)}-1$,
permits one to determine $R(t,r)$ as in Eq. \eqref{eq:evol}.

\end{document}